\documentclass[11pt]{article}
\pdfoutput=1
\usepackage[%
compact=,%
paper=a4paper,%
largepaper=true,%
marginfrac=12,%
runningtitle=false,%
halfparskip,%
font=LatinModern,
biblatex=false,%
natbib=true,%
bibliographystyle=plainurl,%
theorems=numbersfirst,%
theoremswithin=document,%
figure=subfig,%
]{ifiseries}
\usepackage[graphtheory,gametheory,complexitytheory]{newKit}

\newcommand{\nbC}{\mathsf{C}}
\newcommand{\nbG}{\mathsf{G}}
\newcommand{\nbA}{\mathsf{A}}
\newcommand{\nbT}{\mathsf{T}}
\newcommand{\nbN}{\mathsf{N}}

\newcommand{\hc}{\widehat{c}}
\newcommand{\kmer}{\mbox{$k$-mer}\xspace}
\newcommand{\kmers}{\mbox{$k$-mers}\xspace}
\usepackage[draft]{fixme}
\fxsetup{theme=color}
\definecolor{fxtarget}{rgb}{0.8000,0.0000,0.0000}
\usepackage{authblk}

\begin{document}

\title{An Improved Filtering Algorithm\\for Big Read Datasets}
\author[1]{Axel~Wedemeyer\thanks{axw@informatik.uni-kiel.de}}
\author[1]{Lasse~Kliemann\thanks{lki@informatik.uni-kiel.de}}
\author[1]{Anand~Srivastav}
\author[1]{Christian~Schielke}
\author[2]{Thorsten~B.~Reusch}
\author[3]{Philip~Rosenstiel}
\affil[1]{Department of Computer Science\\Kiel~University\\Christian-Albrechts-Platz~4\\24118~Kiel\\Germany}
\affil[2]{Marine~Ecology\\GEOMAR Helmholtz Centre for Ocean Research Kiel\\D\"{u}sternbrooker~Weg~20\\24105~Kiel\\Germany}
\affil[3]{Institute of Clinical Molecular Biology\\Kiel~University\\Schittenhelmstr.~12\\24105~Kiel\\Germany}

\maketitle

\begin{abstract}
\label{sec:abstract}
For single-cell or metagenomic sequencing projects, it is necessary to sequence
with a very high mean coverage in order to make sure that all parts of the
sample DNA get covered by the reads produced.  This leads to huge datasets with
lots of redundant data.  A filtering of this data prior to assembly is
advisable.  Titus Brown et al. (2012) presented the algorithm Diginorm for this
purpose, which filters reads based on the abundance of their $k$-mers.
We present Bignorm, a faster and quality-conscious read filtering algorithm.  An
important new feature is the use of phred quality scores together with a
detailed analysis of the $k$-mer counts to decide which reads to keep.  With
recommended parameters, in terms of median we remove 97.15\% of the reads while
keeping the mean phred score of the filtered dataset high.  Using the SDAdes
assembler, we produce assemblies of high quality from these filtered datasets in
a fraction of the time needed for an assembly from the datasets filtered with
Diginorm.
We conclude that
read filtering is a practical method for reducing read data and for speeding up
the assembly process.  Our Bignorm algorithm allows assemblies of competitive
quality in comparison to Diginorm, while being much faster.  Bignorm is
available for download at
\url{https://git.informatik.uni-kiel.de/axw/Bignorm.git}
\end{abstract}

\section{Background}
\label{sec:background}

Next generation sequencing systems (such as the Illumina platform)
tend to produce an enormous amount of data
--- especially when used for single-cell or metagenomic protocols ---
of which only a small fraction is essential for the assembly of the genome.
It is thus advisable to filter that data prior to assembly.

\subsection{Problem Formulation}\label{sec:problemformulation}

In order to describe our algorithm and its comparison, we need some
formal definitions and concepts.
Denote $\NN \df \set{0,1,2,\hdots}$ the set of non-negative integers,
and for each $n \in \NN$ denote $\setn{n} \df \setft{1}{n}$ the integers from~$1$ to~$n$
(including~$1$ and~$n$).
Denote $\Sig \df \set{\nbA, \nbC, \nbG, \nbT, \nbN}$ the alphabet of nucleotides
plus the symbol $\nbN$ used to indicate an undetermined base.
By $\Sig^*$ we denote all the finite strings over~$\Sig$, and for a $k \in \NN$
by $\Sig^k$ all the strings over $\Sig$ of exactly length~$k$.
For $v \in \Sig^*$, denote $\len{v} \in \NN$ its length and $\bar{v} \in \Sig^*$ its reverse complement.
For $v,w \in \Sig^*$, we write $v \cong w$ if $\len{v}=\len{w}$ and the two strings are equal
up to places where either of them has the $\nbN$ symbol.

The input to the filter algorithm is a \emph{dataset} $D=(n,m,R,Q)$
where for each $i \in \setn{n}$ we have:
\begin{samepage}
  \begin{compactitemize}
  \item $m(i) \in \NN$: a flag for an unpaired ($m(i)=1$) or paired ($m(i)=2$) dataset;
  \item $R(i,s) \in \Sig^*$ for each $s \in \setn{m(i)}$: the set of \emph{reads} in the dataset;
  \item $Q(i,s) \in \ZZ^{\len{R(i,s)}}$ for each $s \in \setn{m(i)}$: the set of corresponding \emph{phred scores}.
  \end{compactitemize}
\end{samepage}
Each read $i \in \setn{n}$ consists of $m(i)$ \emph{read strings} $R(i,1),\hdots,R(i,m(i))$.
For $t \in \setn{\len{R(i,s)}} = \setft{1}{\len{R(i,s)}}$ we denote the nucleotide at position $t$ in read string $R(i,s)$ by $R_t(i,s)$
and its phred score by $Q_t(i,s)$.
Note that in terms of read strings, $D$ may contain the \enquote{same} read multiple times (perhaps with different quality values),
that is, there can be $i \neq j$ such that $R(i) = R(j)$.
Hence it is beneficial that we refer to reads by their indices $1,\hdots,n$.

Denote $x \in \Sig^*$ the genome from which the reads were obtained and $g \df \len{x}$ its length.
(For the purpose of this exposition, we simplify by assuming the genome is a single string.).
For each locus $\ell \in \setn{g}$, the \emph{coverage} $c_\ell(D)$ of $\ell$ with respect to $D$ is
informally described as the number of read strings that were or could have been produced by the sequencing machine while reading
a part of $x$ that contains locus~$\ell$.
More precisely, for each $v \in \Sig^*$ define 
\begin{itemize}
\item $c_\ell(v) \df 1$ if
there is a substring $w$ of $x$ which contains locus $\ell$ and satisfies $v \cong w$ or $v\cong\bar{w}$;
\item $c_\ell(v) \df 0$ otherwise.
\end{itemize}
Then we define:
\begin{equation*}
  c_\ell(D) \df \sum_{i=1}^n \sum_{s=1}^{m(i)} c_\ell(R(i,s))
\end{equation*}

A coverage of $c_\ell(D) \approx 20$ for each $\ell \in \setn{g}$ has been empirically determined as optimal
for a successful assembly of $x$ from~$D$~\cite{2012arXiv1203.4802T}.
On the other hand, in many setups, the coverage for a large number of loci is much higher than~$20$,
often rising up to tens or hundreds of thousands,
especially for single-cell or metagenomic protocols (see \autoref{tab:coverage},
\enquote{max} column for the maximal coverage of the datasets that we use in our experiments).
In order to speed up the assembly process --- or in extreme cases to make it possible in the first place,
given certain restrictions on available RAM and/or time --- a sub-dataset $D'=(n',m',R',Q')$ of $D$ should be determined
such that $n'$ is much smaller than $n$ while not losing essential information.
The goal is that using $D'$, an assembly of similar quality than using $D$ is possible.
We only consider the natural approach to create $D'$ by making a choice for each $i \in \setn{n}$
whether to include read~$i$ in $D'$ or not,
so in particular $(R'(1),\hdots,R'(n'))$ will be a sub-vector of $(R(1),\hdots,R(n))$.
When we include a read in $D'$, we also say that it is \emph{accepted}, whereas when we exclude it, we say it is \emph{rejected}.
On an abstract level, a filtered dataset based on $D$ can be specified by giving a set of indices $A \subseteq \setn{n}$
that consists of exactly the accepted reads.

Many popular assemblers, such as SPAdes~\cite{Bankevich2012}, Platanus~\cite{Kajitani22042014}, or Allpaths-LG~\cite{GnMaPr11},
work with the \emph{de Bruijn graph}, that is based on \kmers.
Fix a parameter $k \in \NN$; typically $21 \leq k$.
The set of \kmers of a string $v \in \Sig^*$, denoted $M(v,k) \subseteq \Sig^k$,
is the set of all strings of length $k$ that are substrings of $v$.
Sometimes we need to consider a \kmer multiple times if it occurs in multiple places in the string,
and the corresponding set is denoted:
\begin{align*}
  M^*(v,k) \df \big\{ (\mu,p) \in \Sig^k \times \NN \suchthat
  \text{$\mu$ is a substring of $v$ starting at position $p$}\big\}
\end{align*}
For a read $i \in \setn{n}$ and read string $s \in \setn{m(i)}$ define $M(i,s,k) \df M(R(i,s),k)$
and $M(i,k) \df \bigcup_{s=1}^{m(i)} M(i,s,k)$, so $M(i,k)$ are all the \kmers that occur in any of the $m(i)$ read strings of~$R(i)$.
Denote also $M^*(i,s,k) \df M^*(R(i,s),k)$.

\subsection{Previous Work}

We briefly survey two prior approaches for read pre-processing, namely \textit{trimming} and \textit{error correction}.
Read trimming programms (see~\cite{10.1371/journal.pone.0085024} for a recent review)
try to cut away the low quality parts of a read (or drop reads whose overall quality is low).
These algorithms can be classified in two groups: \textit{running sum} (Cutadapt, ERNE, SolexaQA with
\lstinline|-bwa| option)~\cite{martin2011cutadapt, Prezza:2012:EAB:2382936.2382938, cox2010solexaqa}
and \textit{window based} (ConDeTri, FASTX, PRINSEQ, Sickle, SolexaQA, and Trimmomatic)~\cite{smeds2011condetri,
  fastx, schmieder2011quality, joshi2011sickle, cox2010solexaqa, Bolger01042014}.
The running sum algorithms take a quality threshold $Q$ as input,
which is subtracted from the phred score of each base of the read.
The algorithms vary in the functions applied to the differences to determine the
quality of a read, the direction in which the read is processed, the function's
quality threshold upon which the cutoff point is determined, and the minimum
length of a read after the cutoff to be accepted.

The window based algorithms on the other hand first cut away the reads's 3' or 5' ends (depending on the algorithm)
whose quality is below a specified minimum quality parameter and then determine a contiguous sequence of high quality
using techniques similar to those used in the running sum algorithms.

All of these trimming algorithms generally work on a per-read basis, reading the
input once and processing only a single read at a time. The drawback of this
approach is that low quality sequences within a read are being dropped even when
these sequences are not covered by any other reads whose quality is high. Also
the phred score of a base is not independent between reads, \ie a base whose
phred score is low in one read is likely to have a low phred score in other
reads as well and thus this low quality segment might get dropped altogether,
creating uncovered regions. On the other hand sequences whose quality and
abundance are high are added over and over although their coverage is already
high enough, which yields higher memory usage than necessary.

Most of the error correction programs (see~\cite{AlRuDoBl16} for a recent review) read the input twice:
a first pass gathers statistics about the data (often \kmer counts)
which in a second pass are used to identify and correct errors.
Some programs trim reads which connot be corrected.
Again, coverage is not a concern: reads which seem to be correct or which can be corrected are always accepted.
According to~\cite{AlRuDoBl16}, the probably best known and most used error correction program is Quake~\cite{KeScSa10}.
Its algorithm is based on two assumptions:
\begin{compactitemize}
\item ``For sufficiently large $k$, almost all single-base errors alter \kmers overlapping the error to versions that do not exist
  in the genome. Therefore, \kmers with low coverage, particularly those occurring just once or twice, usually represent sequencing errors.''
\item Errors follow a Gamma distribution, whereas true \kmers are distributed as per a combination of the Normal and the Zeta distribution.
\end{compactitemize}

In the first pass of the program, a score (based on the phred quality scores of the individual nucleotides)
is computed for each \kmer.
After this, Quake computes a \emph{coverage cutoff} value,
that is, the local minimum of the \kmer spectrum between the Gamma and the Normal maxima.
All \kmers having a score higher than the coverage cutoff are considered to be correct
(\emph{trusted} or \emph{solid} in error correction terminology), the others are assumed to be erroneous.
In a second pass, Quake reads the input again and tries to replace erroneous \kmers by trusted ones using a maximum likelihood approach.
Reads which cannot be corrected are optionally trimmed or dumped.

But the main goal of error correctors is not the reduction of the data volume
(in particular, they do not pay attention to excessive coverage),
hence they cannot replace the following approaches.

Titus Brown \emph{et al.}\@ invented an algorithm named \emph{Diginorm}~\cite{2012arXiv1203.4802T,10.1371/journal.pone.0101271}
for read filtering that rejects or accepts reads based on the abundance of their \kmers.
The name \emph{Diginorm} is a short form for \emph{digital normalization}:
the goal is to normalize the coverage over all loci, using a computer algorithm after sequencing.
The idea is to reject those reads which mainly bring \kmers that have been seen many times in other reads already.
Diginorm processes reads one by one.
Let the read currently processed be~$i \in \setn{n}$.
For each \kmer $\mu \in \Sig^k$, define
\begin{align*}
  c(\mu, i) \df \big\lvert\big\{ j \in\NN \suchthat (j < i) \text{ and }(\mu \in M(j, k))
  \text{ and (read $j$ was previously accepted)} \big\}\big\rvert \comma
\end{align*}
which says in how many accepted reads we have seen the \kmer $\mu$ so far.
In order to save RAM, Diginorm does not keep track of those numbers exactly,
but instead keeps appropriate estimates $\hc(\mu, i)$ using the count-min sketch (CMS)~\cite{CM05}.
For each $i \in \setn{n}$ and $s \in \setn{m(i)}$ denote the vector $C(i,s) \df \parens{\hc(\mu,i)}_{\mu\in M(i,s,k)}$.
The read~$i$ is accepted if the median of the numbers in $C(i,s)$ is below a fixed threshold, usually~$20$,
for each $s \in \setn{m(i)}$.
It was demonstrated that successful assemblies are still possible after Diginorm removed the majority of the data.

\subsection{Our Algorithm}\label{sec:ouralgorithm}

Diginorm is a pioneering work. However, the following points,
which are important from the biological or computational quality
point of view, are not covered in Diginorm. We present them as an enhancement
in our work:

\begin{enumerate}\label{enhancements}
\item[(i)] We incorporate the important phred quality score
  into the decision whether to accept or to reject a read,
  using a quality threshold.
  This allows a tuning of the filtering process towards high\-/quality assemblies,
  by using different thresholds.
\item[(ii)]
  When deciding whether to accept or to reject read~$i$,
  we do a detailed analysis of the numbers in the vectors $C(i,s)$.
  Diginorm merely considers their medians.
\item[(iii)]
  We offer a better handling of the $\nbN$ case, that is, when the sequencing machine could
  not decide for a particular nucleotide.
  Diginorm simply converts all $\nbN$ to $\nbA$, which can lead to false \kmer counts.\footnote{%
    We have observed some evidence that this may lead to a spuriously higher GC content.
    This will be investigated in future work.}
\item[(iv)]
  We provide a substantially faster implementation.
  For example, we include fast hashing functions (see~\cite{DIETZFELBINGER199719,Wol03})
  for counting \kmers through the count-min sketch data structure (CMS),
  and we use the C programming language and OpenMP.
\end{enumerate}

A detailed description of our algorithm, called \emph{Bignorm}, is given in the
next section.  Its name was chosen to emphasize the goal of drastically reducing
massive datasets.

Bignorm, like Diginorm, is based on the count-min sketch (CMS) for counting \kmers.
CMS is a probabilistic data structure for counting objects from a large universe.
We give a brief and abstract description.
Let $a = (\eli{a}{N}) \in \NN^N$ be a vector, given implicitly as a sequence of updates of the form $(p,\Del)$
with $p \in \setn{N}$ and $\Del \in \NN$.
Each update $(p,\Del)$ modifies $a$ in the way $a_p \df a_p + \Del$; where initially $a=(0,\hdots,0)$.
If $\Del=1$ in each update, then an interpretation of the vector $a$ is that we count how many times
we observe each of the objects identified by the numbers in $\setn{N}$.
If $N$ is large, \eg if $N$ is the number $4^k$ of all possible \kmers (we do not count \kmers with $\nbN$ symbols),
then we may not be able to store $a$ in RAM.\
(For example, the typical choice of $k=21$ brings $a$ into terabyte range;
in our experiments we use $k=32$.)
Instead we fix two parameters: the \emph{width} $m \in \NN$ and the \emph{depth} $t \in \NN$
and store a matrix of $m \cdot t$ \emph{CMS counters} $c_{p,q}$ with $p \in \setn{m}$ and $q \in \setn{t}$.
Moreover, we randomly draw $t$ hash functions $\eli{h}{t}$ from a universal family.
Each $h_q$ maps from $\setn{N}$ to~$\setn{m}$.
Initially, all counters in the matrix are zero.
Upon arrival of an update $(p,\Del)$, for each row $q \in \setn{t}$ we update $c_{h_q(p), q} \df c_{h_q(p), q} + \Del$.
That is, for each row $q$ we use the hash function $h_q$ to map from the larger space $\setn{N}$
(from which the index $p$ comes) to the smaller space $\setn{m}$
of possible positions in the row.
Denote
\begin{equation}
  \label{cms-estimate}
  \widehat{a}_p \df \min \set{c_{h_1(p), 1}, \hdots, c_{h_t(p), t}} \period
\end{equation}
Then it can be proved~\cite{CM05} that $\widehat{a}_p$ is an estimate of $a_p$ in the following sense:
clearly $a_p \leq \widehat{a}_p$, and
with probability at least $1 - e^{1-t}$ we have $\widehat{a}_p \leq \frac{e}{m-1} \sum_{j=1}^N a_j$.
The probability is over the choice of hash functions.
For example, choosing $t \df 10$ is enough to push the error probability, upper-bounded by $e^{1-t}$, below $0.013\%$.

In our application, $N=4^k$ is the number of possible \kmers (without $\nbN$ symbols)
and we implement a bijection $\fn{\bet}{\Sig^k \map \setn{N}}$,
so we can identify each \kmer $\mu$ by a number $\bet(\mu) \in \setn{N}$.
Upon accepting some read $i$, we update the CMS counters using all the updates of the form $(\bet(\mu), 1)$
with $\mu \in M(i,k)$ not containing the $\nbN$ symbol,
that is, for each such $\mu$ we increase the count $\bet(\mu)$ by $\Del=1$.
Then when all the reads $1,\hdots,i-1$ have been processed,
the required count $c(\mu,i)$ corresponds to the entry $a_{\bet(\mu)}$ in the vector $a$ used in the description of CMS,
and for the estimate $\hc(\mu,i)$ we can use the estimate $\widehat{a}_{\bet(\mu)}$ as given in~\eqref{cms-estimate}.

\section{Methods}

\subsection{Description of Bignorm}

We give a detailed description of our enhancements (i) to (iv) that were briefly lined out \vpageref{enhancements}.
Although most of the settings are generic, in some places we assume that data comes from the Illumina.

We start with (i), (ii), and (iii).
Fix a read $i \in \setn{n}$ and a read string $s \in \setn{m(i)}$.
Recall that for each $t \in \setn{\,\len{R(i,s)}\,}$ the nucleotide  $R_t(i,s)$ at position $t$ in the read string $R(i,s)$
is associated with a quality value $Q_t(i,s)$ known as \emph{phred score}.
We want to assign a single value $Q(i,s,\mu,p)$ to each $(\mu,p) \in M^*(i, s, k)$.
We do so by taking the minimum phred score over the nucleotides in $\mu$ when aligned at position $p$, that is:
\begin{equation*}
  Q(i,s,\mu,p) \df \min_{t=p}^{p+k-1} Q_t(i,s)
\end{equation*}
($\mu$ occurs on the right-hand side only implicitely through its length $k$.)

Fix the following parameters:
\begin{compactitemize}
\item \emph{$\nbN$-count threshold} $N_0 \in \NN$, which is $10$ by default;
\item \emph{quality threshold} $Q_0 \in \ZZ$, which is $20$ by default;
\item \emph{rarity threshold} $c_0 \in \NN$, which is $3$ by default;
\item \emph{abundance threshold} $c_1 \in \NN$, which is $20$ by default;
\item \emph{contribution threshold} $B \in \NN$, which is $3$ by default.
\end{compactitemize}
When our algorithm has to decide whether to accept or to reject a read $i \in \setn{n}$, it performs the following steps.
If the number of $\nbN$ symbols counted over all $m(i)$ read strings in $i$ is larger than $N_0$,
the read is rejected right away.
Otherwise, for each $s \in \setn{m(i)}$ define the set of \emph{high-quality \kmers}:
\begin{align*}
  H(s) \df \big\{ (\mu,p) \in M^*(i,s,k) \suchthat (Q_0 \leq Q(i,s,\mu,p))
  \text{ and ($\mu$ does not contain $\nbN$)} \big\}
\end{align*}

We determine the \emph{contribution} of $R(i,s)$ to \kmers of different frequencies:
\begin{align*}
  b_0(s) & \df \card{\set{ (\mu,p) \in H(s) \suchthat \hc(\mu, i) < c_0 }} \\
  b_1(s) & \df \card{\set{ (\mu,p) \in H(s) \suchthat c_0 \leq \hc(\mu, i) < c_1 }}
\end{align*}
Note that the frequencies are determined via CMS counters and do not consider the position $p$ at which the \kmer is found in the read string.
The read $i$ is accepted if and only if at least one of the following conditions is met:
\begin{align}
  \label{accept1}
 b_0(s) & > k\text{ for at least one read string $s$}\\
  \label{accept2}
  \sum_{s=1}^{m(i)} b_1(s) & \geq B
\end{align}
If the read is accepted, then for each $\mu \in M(i,k)$ the corresponding CMS counter is incremented,
provided that $\mu$ does not contain the $\nbN$ symbol.
Then processing of the next read starts.

The rationale for condition~\eqref{accept1} is as follows.
If a \kmer is seen less than $c_0$ times, we suspect it to be the result of a read error.
However, if more than $k$ \kmers in a read string contain an error, this read string must have more than one erroneous nucleotide.
This is not likely for the Illumina platform, since there, most errors are single substitutions~\cite{KeScSa10}.
So if $b_0(s) > k$ for some $s$, then the read string $R(i,s)$ should be assumed to correctly contain a \emphasis{rare} \kmer,
so it must not be filtered out.

Condition~\eqref{accept2} says that in the read $i$, there are enough (namely at least $B$) \kmers
where each of them is too frequent to be a read error (CMS counters at least $c_0$)
but not so abundant that it should be considered redundant (CMS counters less than $c_1$).

This concludes the description of (i), (ii), and (iii),
namely how we analyze the counts in $C(i,s) = \parens{\hc(\mu,i)}_{\mu\in M(i,s,k)}$ for each read $i$ and $s \in \setn{m(i)}$,
how we incorporate quality information, and how we handle the $\nbN$ symbol.

Finally, to accomplish (iv),
we wrote a multi-threaded implementation completely in the C programming language.
The parallel code uses OpenMP.\
For comparison, the implementation of the original Diginorm algorithm (included in the khmer-package~\cite{khmer2014})
features a single-threaded design and is written in Python and \CXX;\@
strings have to be converted between Python and \CXX\ at least twice.

\subsection{Experimental Setup}

For the experimental evaluation, we collected the following datasets.
We use two single cell datasets of the UC San Diego, one of the group of Ute Hentschel (now GEOMAR Kiel)
and 10 datasets from the JGI Genome Portal.
The datasets from JGI were selected as follows.
On the JGI Genome Portal~\cite{jgi}, we used \enquote{single cell} as search term.
We narrowed the results down to datasets which had all of the following properties:
\begin{compactitemize}
\item status \enquote{complete};
\item containing read data \emphasis{and} an assembly in the download section;
\item aligning the reads to the assembly using bowtie2~\cite{Langmead2012} yields an \enquote{overall alignment rate} of more than~$70\%$.
\end{compactitemize}
From those datasets, we arbitrarily selected one per species, until we had a collection of $10$ datasets.
We refer to each combination of species and selected dataset as a \emph{case} in the following.
In total, we have $13$ cases; the details are given in \autoref{tab:cases}.

\begin{table}[h!]
  \centering
  \small
  \begin{tabular}{@{}lllr@{}}\toprule
    Short Name  & Species/Description                                   & Source                                 & URL                   \\\midrule
    ASZN2       & Candidatus Poribacteria sp. WGA-4E\_FD                & Hentschel Group~\cite{kamke2013single} & \cite{hentschelWGA4E} \\
    Aceto       & Acetothermia bacterium JGI MDM2 LHC4sed-1-H19         & JGI Genome Portal                      & \cite{Aceto}          \\
    Alphaproteo & Alphaproteobacteria bacterium SCGC AC-312\_D23v2      & JGI Genome Portal                      & \cite{Alphaproteo}    \\
    Arco        & Arcobacter sp. SCGC AAA036-D18                        & JGI Genome Portal                      & \cite{Arco}           \\
    Arma        & Armatimonadetes bacterium JGI 0000077-K19             & JGI Genome Portal                      & \cite{Arma}           \\
    Bacteroides & Bacteroidetes bacVI JGI MCM14ME016                    & JGI Genome Portal                      & \cite{Bacteroidetes}  \\
    Caldi       & Calescamantes bacterium JGI MDM2 SSWTFF-3-M19         & JGI Genome Portal                      & \cite{Caldi}          \\
    Caulo       & Caulobacter bacterium JGI SC39-H11                    & JGI Genome Portal                      & \cite{Caulo}          \\
    Chloroflexi & Chloroflexi bacterium SCGC AAA257-O03                 & JGI Genome Portal                      & \cite{Chloroflexi}    \\
    Crenarch    & Crenarchaeota archaeon SCGC AAA261-F05                & JGI Genome Portal                      & \cite{Crenarch}       \\
    Cyanobact   & Cyanobacteria bacterium SCGC JGI 014-E08              & JGI Genome Portal                      & \cite{Cyanobact}      \\
    E.coli      & E.coli K-12, strain MG1655, single cell MDA, Cell one & UC San Diego                           & \cite{ucsdsinglecell} \\
    SAR324      & SAR324 (Deltaproteobacteria)                          & UC San Diego                           & \cite{ucsdsinglecell} \\\bottomrule
  \end{tabular}
  \caption{Selected Species and Datasets (Cases)\label{tab:cases}}
\end{table}

For each case, we analyze the results obtained with Diginorm
and with Bignorm using quality parameters $Q_0 \in \set{ 5, 8, 10, 12, 15, 18, 20, \hdots, 45}$.
Analysis is done on the one hand in terms of data reduction, quality, and coverage.
On the other hand, we study actual assemblies that are computed with SPAdes~\cite{Bankevich2012}
based on the raw and filtered datasets.
All the details are given in the next section.

The dimensions of the count-min sketch are fixed to $m=1024$ and $t=10$, thus $10$~GB of RAM where used.

\section{Results}

We do analysis in large parts by looking at percentiles and quartiles.
The $i$th quartile is denoted Q$i$, where we use Q0 for the minimum,
Q2 for the median, and Q4 for the maximum.
The $i$th percentile is denoted~P$i$; we often use the $10$th percentile~P10.

\subsection{Number of Accepted Reads}

\begin{figure}[t]
  \centering
  \subfloat[][Percentage of accepted reads (\ie reads kept) over all datasets.\label{fig:readskept}]
  {\includegraphics[width=0.49\textwidth]{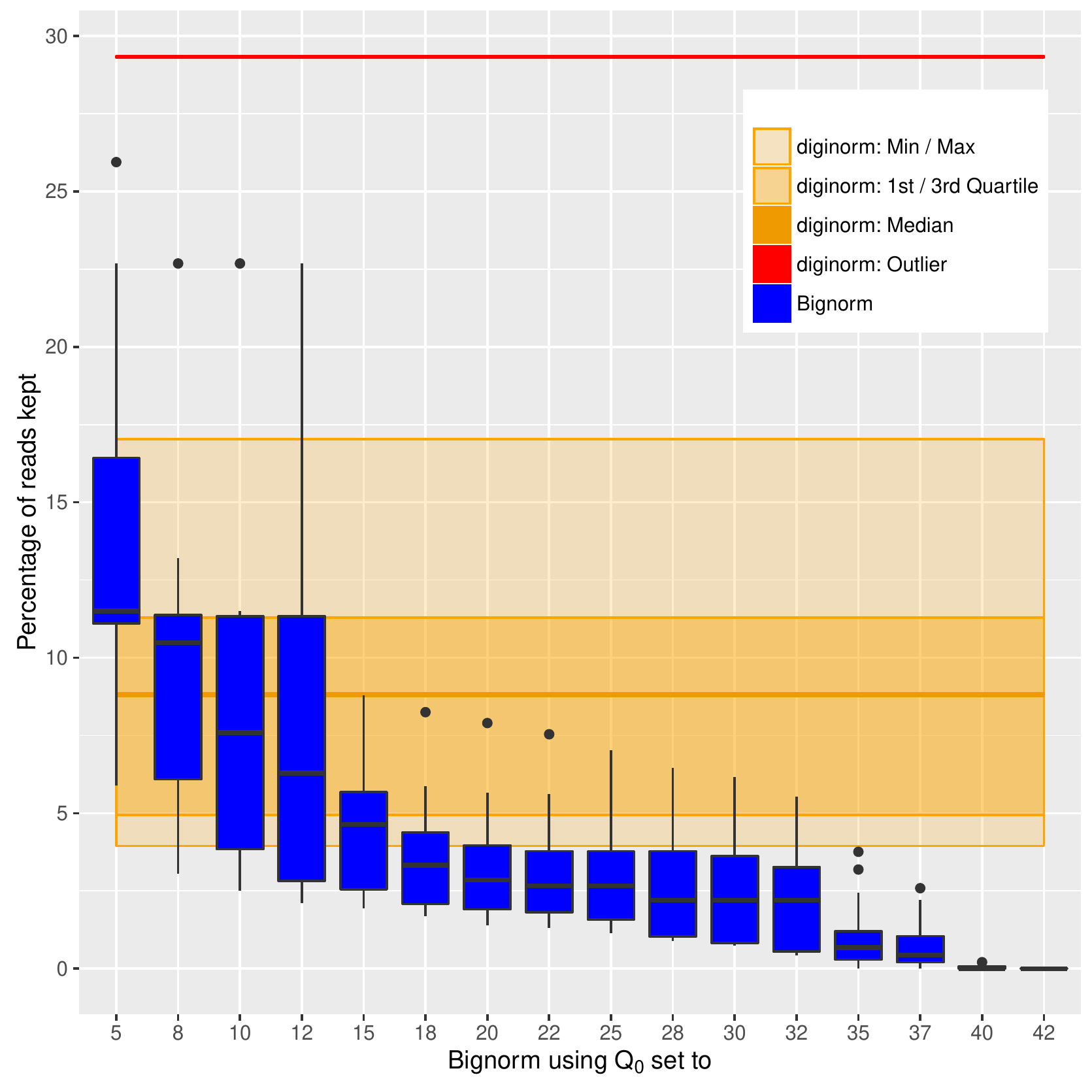}}
  \hfill
  \subfloat[][Mean quality values of the accepted reads over all datasets.\label{fig:quality}]
  {\includegraphics[width=0.49\textwidth]{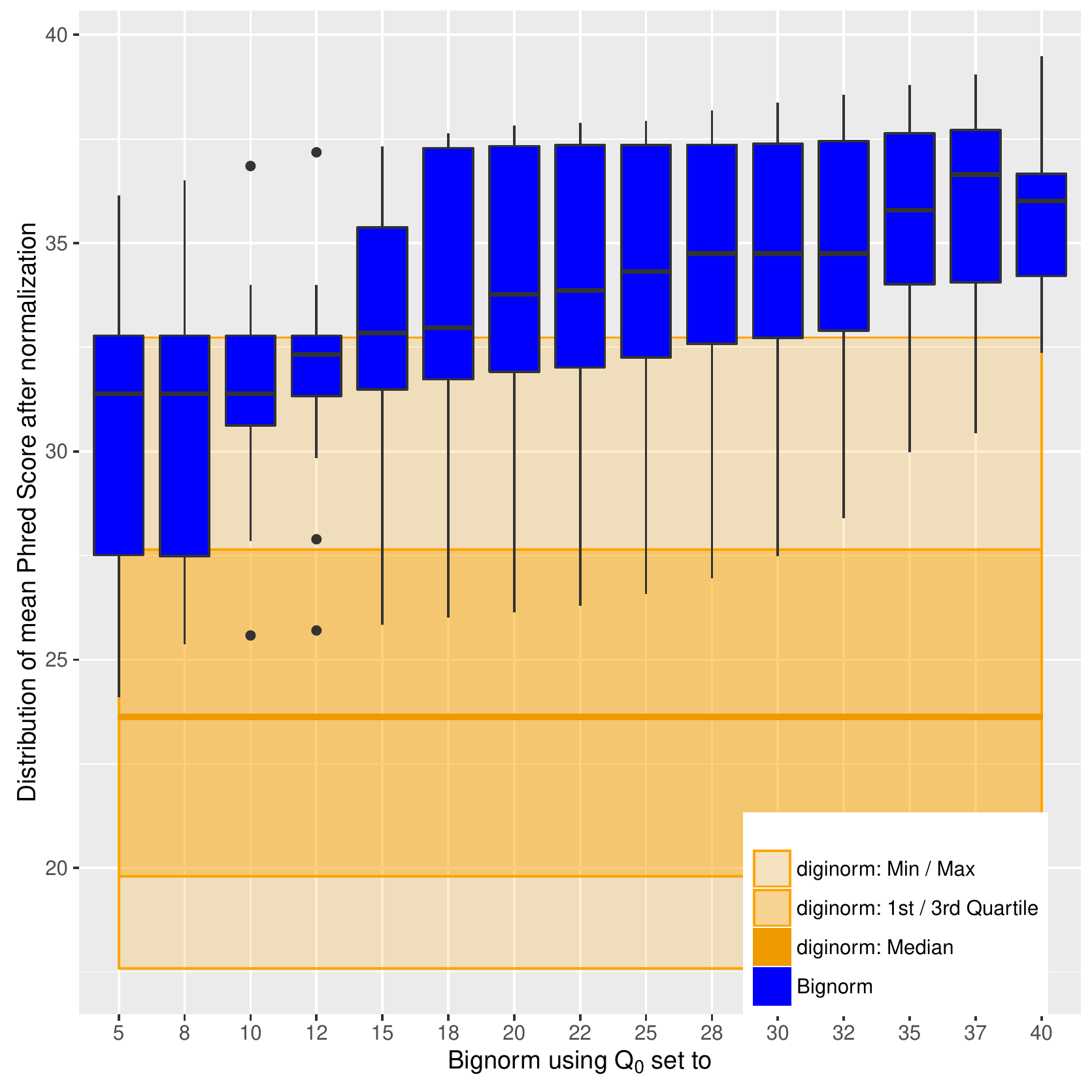}}
  \caption{Boxplots showing reduction and quality statistics.}
\end{figure}

Statistics for the number of accepted reads are given as a boxplot in \autoref{fig:readskept} \vpageref{fig:readskept}.
This plot is constructed as follows.
Each of the blue boxes corresponds to Bignorm with a particular $Q_0$,
while Diginorm is represented as the wide orange box in the background (recall that Diginorm does not consider quality values).
Note that the \enquote{whiskers} of Diginorm's box are shown as light\-/orange areas.
For each box, for each case the raw dataset is filtered using the algorithm and algorithmic parameters corresponding to that box,
and the percentage of the accepted reads is taken into consideration.
So for example, if the top of a box (which corresponts to the 3rd quartile, also denoted Q3) gives the value $x \%$,
then we know that for $75\%$ of the cases, $x \%$ or less of the reads were accepted
using the algorithm and algorithmic parameters corresponding to the box.

There are two prominent outliers: one for Diginorm with value $\approx 29\%$ (shown as the red line at the top)
and one for Bignorm for $Q_0=5$ with value $\approx 26\%$.
In both cases the Arma dataset is responsible, for which we do not have an explanation at this time.
For $15 \leq Q_0$, even Bignorm's outliers fall below Diginorm's median,
and for $18 \leq Q_0$ Bignorm keeps less than $5\%$ of the reads for at least $75\%$ of the datasets.
In the range $20 \leq Q_0 \leq 25$, Bignorm delivers similar results for the different $Q_0$,
and the gain in reduction for larger $Q_0$ is small up to $Q_0 = 32$.
For even larger $Q_0$, there is another jump in reduction, but we will see that coverage
and the quality of the assembly suffer too much in that range.
We conjecture that in the range $18 \leq Q_0 \leq 32$, we remove most of the actual errors,
whereas for larger $Q_0$ we also remove useful information.

\subsection{Quality Values}

Statistics for phred quality scores in the filtered datasets are given in \autoref{fig:quality} \vpageref{fig:quality}.
The data was obtained using \lstinline|fastx_quality_stats| from the FASTX~Toolkit~\cite{fastx} on the filtered fastq files and
calculating the mean phred quality scores over all read positons for each dataset. Looking at the statistics for these overall means,
for $15 \leq Q_0$, Bignorm's median is better than Diginorm's maximum.
For $20 \leq Q_0$, this effect becomes even stronger.
For all $Q_0$ values, Bignorm's minimum is clearly above Diginorm's median.
Note that $10$ units more means reducing error probability by factor~$10$.

In \autoref{tab:quality}, we give quartiles of mean quality values for the raw datasets and Bignorm's datasets produced with $Q_0 = 20$.
Bignorm improves slightly on the raw dataset in all five quartiles.

\begin{table}[h!]
  \centering
  \begin{tabular}{lrr} \toprule
    Quartile    & Bignorm & raw   \\ \midrule
    Q4 (max)    & 37.82   & 37.37 \\
    Q3          & 37.33   & 36.52 \\
    Q2 (median) & 33.77   & 32.52 \\
    Q1          & 31.91   & 30.50 \\
    Q0 (min)    & 26.14   & 24.34 \\ \bottomrule
  \end{tabular}
  \caption{Comparing quality values for the raw dataset and Bignorm with
    $Q_0=20$.\label{tab:quality}}
\end{table}

Of course, all this could be explained by Bignorm simply cutting away any low-quality reads.
However, the data in the next section suggests that Bignorm may in fact be more careful than this.

\subsection{Coverage}

\begin{figure}[t]
  \centering
  \subfloat[][Mean coverage over all datasets.\label{fig:coverage:1}]{\includegraphics[width=0.49\textwidth]{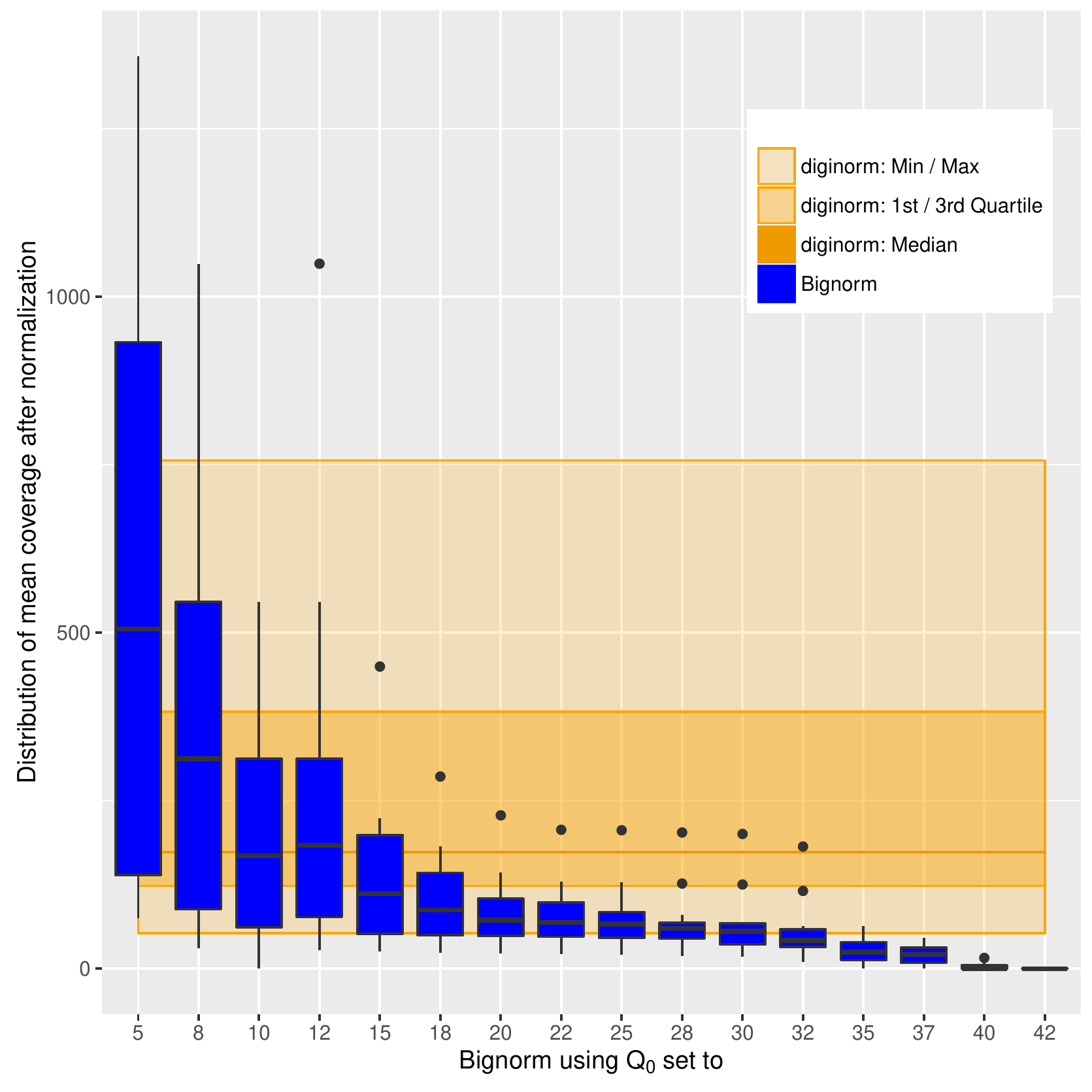}}
  \hfill
  \subfloat[][10th percentile of the coverage over all datasets.\label{fig:coverage:2}]{\includegraphics[width=0.49\textwidth]{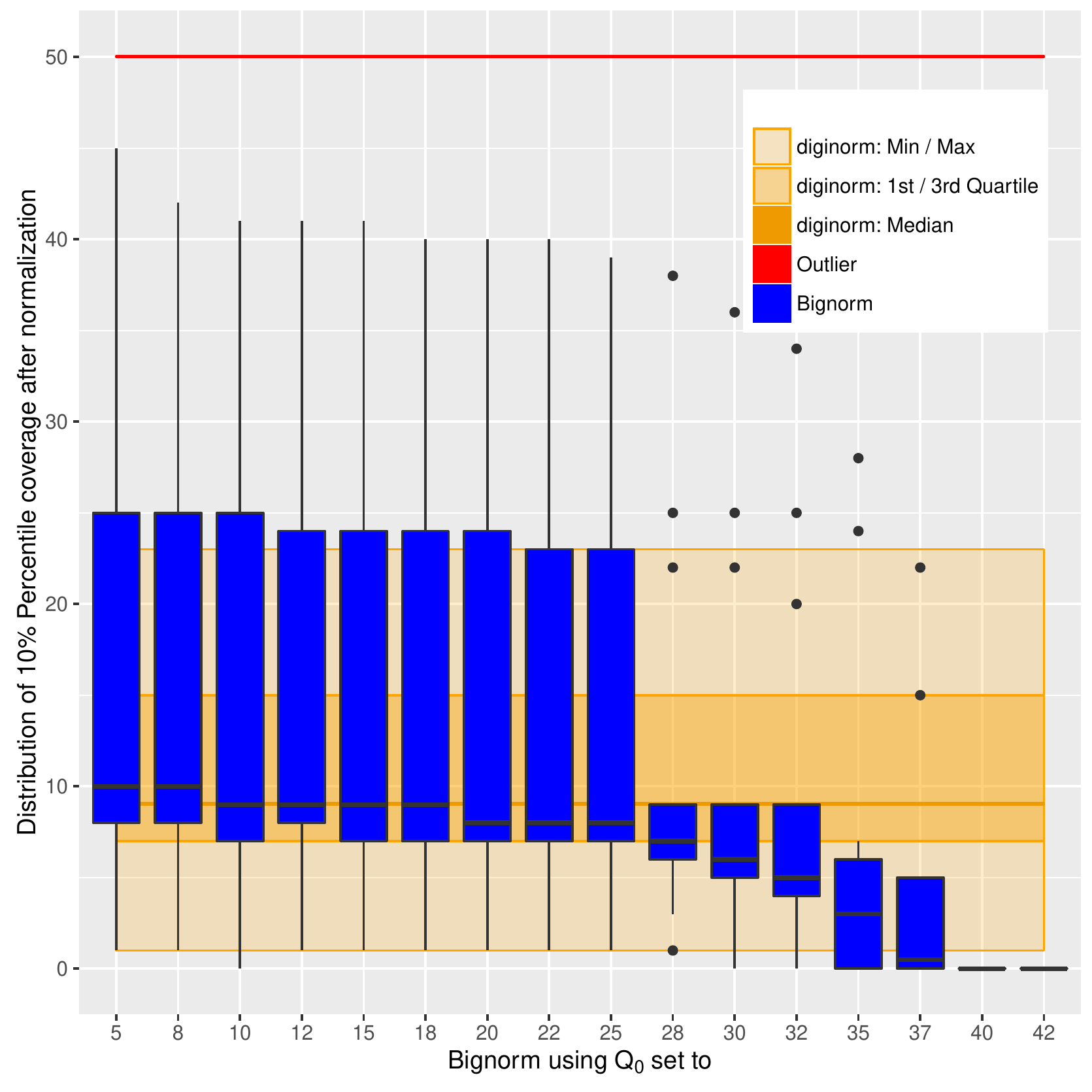}}
  \caption{Boxplots showing coverage statistics.\label{fig:coverage}}
\end{figure}

In \autoref{fig:coverage} \vpageref{fig:coverage} we see statistics for the coverage.
The data was obtained by remapping the filtered reads onto the assembly from the JGI using bowtie2
and then using \lstinline|coverageBed| from the bedtools~\cite{Quinlan15032010} and R~\cite{R} for the statistics.
In \autoref{fig:coverage:1}, the mean is considered.
For $15 \leq Q_0$, Bignorm reduces the coverage heavily.
For $20 \leq Q_0$, Bignorm's Q3 is below Diginorm's Q1.
This may raise the concern that Bignorm could create areas with insufficient coverage.
However, in \autoref{fig:coverage:2}, we look at the 10th percentile (P10) of the coverage instead of the mean.
We consider this statistics as an indicator for the impact of the filtering on areas with low coverage.
For $Q_0 \leq 25$, Bignorm's Q3 is on or above Diginorm's maximum,
and Bignorm's minimum coincides with Diginorm's (except for $Q_0=10$, where we are slightly below).
In terms of median, both algorithms are very similar for $Q_0 \leq 25$.
We consider all this as a strong indication that we cut away in the right places.

For $28 \leq Q_0$, there is a clear drop in coverage, so we do not recommend such $Q_0$ values.

In \autoref{tab:coverage}, we give coverage statistics for each dataset.
The reduction compared to the raw dataset in terms of mean, P90, and maximum is substantial.
But also the improvement of Bignorm over Diginorm in mean, P90, and maximum is considerable for most datasets.

\begin{table}
  \begin{center}
    \small
    \begin{tabular}{llrrrr}\toprule
  Dataset & Algorithm & P10 & $\mathrm{mean}$ & P90 & $\max$ \\
  \midrule
  \multirow{3}{*}{Aceto} & Bignorm &   6 & 132 & 216 & 6801 \\
          & Diginorm &   7 & 171 & 295 & 12020 \\
          & raw &  15 & 9562 & 17227 & 551000 \\
  \midrule
  \multirow{3}{*}{Alphaproteo} & Bignorm &  10 &  43 &  92 & 884 \\
          & Diginorm &   7 & 173 & 481 & 6681 \\
          & raw &  25 & 5302 & 14070 & 303200 \\
  \midrule
  \multirow{3}{*}{Arco} & Bignorm &   1 &  98 &  54 & 2103 \\
          & Diginorm &   1 & 362 & 200 & 6114 \\
          & raw &   3 & 10850 & 4091 & 220600 \\
  \midrule
  \multirow{3}{*}{Arma} & Bignorm &   8 &  23 &  32 & 358 \\
          & Diginorm &   8 &  79 & 141 & 5000 \\
          & raw &  17 & 629 & 1118 & 31260 \\
  \midrule
  \multirow{3}{*}{ASZN2} & Bignorm &  40 &  70 &  83 & 2012 \\
          & Diginorm &  23 & 143 & 354 & 3437 \\
          & raw &  50 & 1738 & 4784 & 43840 \\
  \midrule
  \multirow{3}{*}{Bacteroides} & Bignorm &   3 &  74 &  90 & 6768 \\
          & Diginorm &   3 & 123 & 205 & 7933 \\
          & raw &   7 & 6051 & 8127 & 570900 \\
  \midrule
  \multirow{3}{*}{Caldi} & Bignorm &  25 &  63 & 110 & 786 \\
          & Diginorm &  15 &  67 & 135 & 3584 \\
          & raw &  27 & 1556 & 3643 & 33530 \\
  \midrule
  \multirow{3}{*}{Caulo} & Bignorm &   7 & 228 & 216 & 10400 \\
          & Diginorm &   8 & 362 & 491 & 35520 \\
          & raw &   8 & 10220 & 9737 & 464300 \\
  \midrule
  \multirow{3}{*}{Chloroflexi} & Bignorm &   8 &  72 & 101 & 2822 \\
          & Diginorm &   9 & 412 & 878 & 20850 \\
          & raw &   9 & 5612 & 7741 & 316900 \\
  \midrule
  \multirow{3}{*}{Crenarch} & Bignorm &   8 & 104 & 159 & 3770 \\
          & Diginorm &  10 & 560 & 1285 & 29720 \\
          & raw &  10 & 8086 & 14987 & 316700 \\
  \midrule
  \multirow{3}{*}{Cyanobact} & Bignorm &   9 & 144 & 153 & 5234 \\
          & Diginorm &  10 & 756 & 1450 & 26980 \\
          & raw &  10 & 9478 & 11076 & 356600 \\
  \midrule
  \multirow{3}{*}{E.coli} & Bignorm &  37 &  45 &  56 & 234 \\
          & Diginorm &  50 & 382 & 922 & 7864 \\
          & raw & 112 & 2522 & 6378 & 56520 \\
  \midrule
  \multirow{3}{*}{SAR324} & Bignorm &  24 &  49 &  71 & 1410 \\
          & Diginorm &  18 &  53 & 107 & 2473 \\
          & raw &  26 & 1086 & 2761 & 106000 \\
  \bottomrule
\end{tabular}

    \caption{Coverage statistics for Bignorm with $Q_0 = 20$, Diginorm, and the
      raw datasets.\label{tab:coverage}}
  \end{center}
\end{table}

\subsection{Assesment through Assemblies}

\begin{figure}[t]
  \centering
  \includegraphics[width=\textwidth]{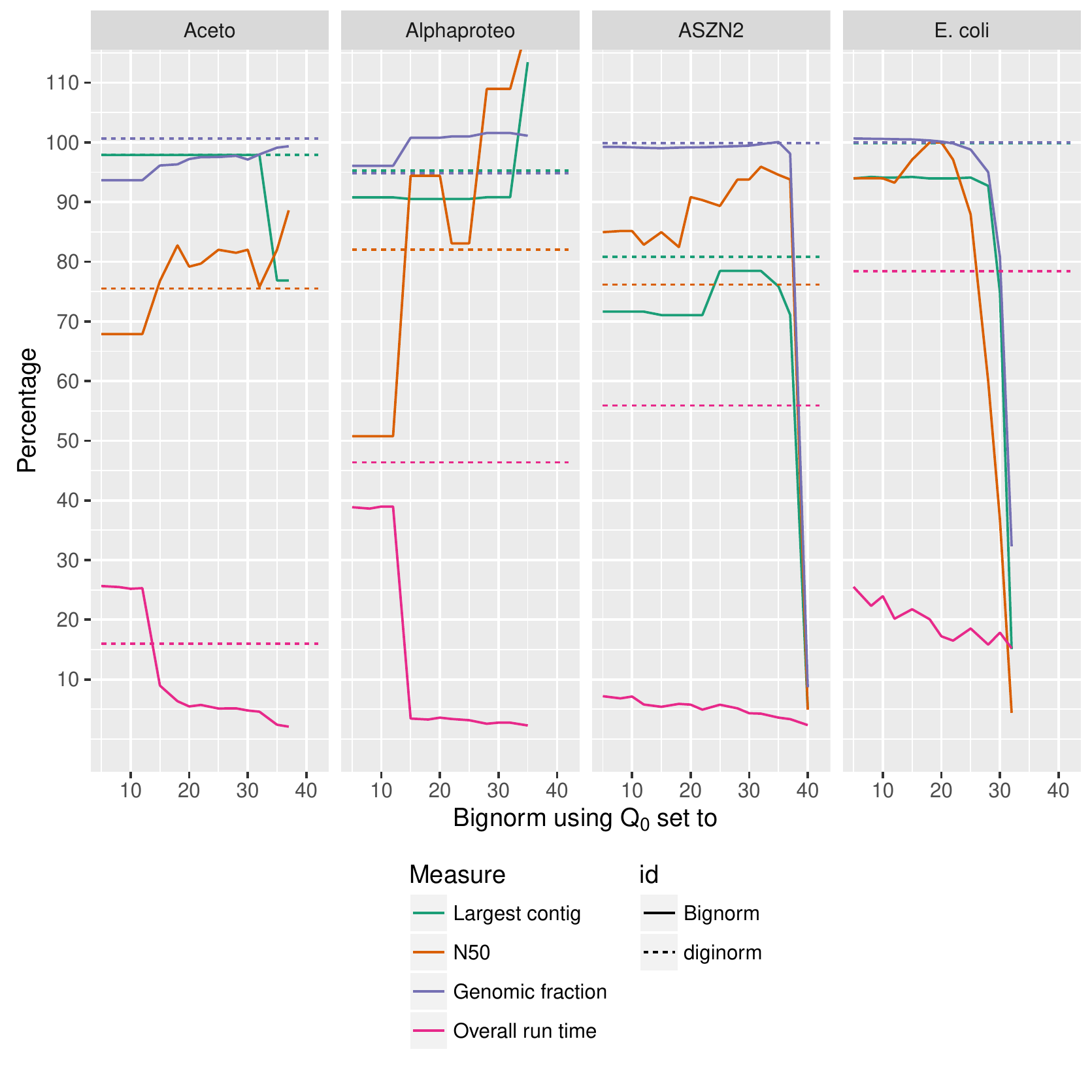}
  \caption{Statistics for the assemblies of four selected datasets.}
\label{fig:quast}
\end{figure}

The quality and significance of read filtering
is subject to complete assemblies, which is the final \enquote{road test}
of  algorithms.
For each case, we do an assembly with SPAdes
using the raw dataset and those filtered with Diginorm and Bignorm for a selection of $Q_0$ values.
The assemblies are then analyzed using quast~\cite{Gurevich15042013} and the assembly from the JGI as reference.
Statistics for four cases are shown in \autoref{fig:quast}.
We give the quality measures N50, genomic fraction, and largest contig,
and in addition the overall running time (pre-processing plus assembly).
Each measure is given in percent relative to the raw dataset.

Generally, our biggest improvements are for N50 and running time.
For $15 \leq Q_0$, Bignorm is always faster than Diginorm, for three of the four cases by a large margin.
In terms of N50, for $15 \leq Q_0$ we observe improvements for three cases.
For E.coli, Diginorm's N50 is $100\%$, that we also attain for $Q_0 = 20$.
In terms of genomic fraction and largest contig, we cannot always attain the same quality as Diginorm;
the biggest deviation at $Q_0=20$ is $10$ percentage points for the ASZN2 case.
The N50 is generally accepted as one of the most important measures,
as long as the assembly respresents the genome well (as mesured here by the genomic fraction)~\cite{Earl01122011}.

\begin{table}
  \begin{center}
    \footnotesize
    \begin{tabular}{llrrrrr}\toprule
 \multirow{2}{*}{Dataset}    & \multirow{2}{*}{Algorithm} & reads kept & mean phred & contigs                    & filter time & SPAdes time \\
                             &                            & in \%      & score      & \scriptsize{$\geq10\,000$} & in sec      & in sec      \\\midrule
\multirow{3}{*}{Aceto}       & Bignorm                    & 3.16       & 37.33      & 1                          & 906         & 1708        \\
                             & Diginorm                   & 3.95       & 27.28      & 1                          & 3290        & 4363        \\
                             & raw                        &            & 36.52      & 3                          &             & 47813       \\ \midrule
\multirow{3}{*}{Alphaproteo} & Bignorm                    & 3.13       & 34.65      & 18                         & 623         & 420         \\
                             & Diginorm                   & 7.81       & 28.73      & 17                         & 1629        & 11844       \\
                             & raw                        &            & 33.64      & 17                         &             & 29057       \\ \midrule
\multirow{3}{*}{Arco}        & Bignorm                    & 2.20       & 33.77      & 4                          & 429         & 207         \\
                             & Diginorm                   & 8.76       & 21.39      & 6                          & 1410        & 1385        \\
                             & raw                        &            & 32.27      & 6                          &             & 15776       \\ \midrule
\multirow{3}{*}{Arma}        & Bignorm                    & 7.90       & 28.21      & 44                         & 240         & 135         \\
                             & Diginorm                   & 29.30      & 21.19      & 50                         & 588         & 1743        \\
                             & raw                        &            & 26.96      & 44                         &             & 5371        \\ \midrule
\multirow{3}{*}{ASZN2}       & Bignorm                    & 5.66       & 37.66      & 118                        & 1224        & 1537        \\
                             & Diginorm                   & 12.62      & 32.73      & 130                        & 5125        & 21626       \\
                             & raw                        &            & 36.85      & 112                        &             & 47859       \\ \midrule
\multirow{3}{*}{Bacteroides} & Bignorm                    & 2.85       & 37.47      & 6                          & 653         & 3217        \\
                             & Diginorm                   & 4.94       & 27.64      & 5                          & 2124        & 3668        \\
                             & raw                        &            & 37.25      & 9                          &             & 32409       \\ \midrule
\multirow{3}{*}{Caldi}       & Bignorm                    & 3.97       & 37.82      & 41                         & 842         & 455         \\
                             & Diginorm                   & 5.61       & 30.67      & 36                         & 1838        & 793         \\
                             & raw                        &            & 37.37      & 38                         &             & 7563        \\ \midrule
\multirow{3}{*}{Caulo}       & Bignorm                    & 2.40       & 36.95      & 10                         & 679         & 712         \\
                             & Diginorm                   & 4.70       & 25.16      & 9                          & 2584        & 765         \\
                             & raw                        &            & 36.01      & 13                         &             & 18497       \\ \midrule
\multirow{3}{*}{Chloroflexi} & Bignorm                    & 1.40       & 31.91      & 32                         & 694         & 134         \\
                             & Diginorm                   & 9.70       & 18.91      & 33                         & 2304        & 1852        \\
                             & raw                        &            & 30.50      & 34                         &             & 15108       \\ \midrule
\multirow{3}{*}{Crenarch}    & Bignorm                    & 1.46       & 33.18      & 19                         & 1107        & 790         \\
                             & Diginorm                   & 9.72       & 19.80      & 18                         & 2931        & 3754        \\
                             & raw                        &            & 31.49      & 26                         &             & 20590       \\ \midrule
\multirow{3}{*}{Cyanobact}   & Bignorm                    & 1.65       & 30.45      & 12                         & 679         & 450         \\
                             & Diginorm                   & 11.30      & 17.58      & 13                         & 1487        & 1343        \\
                             & raw                        &            & 28.49      & 13                         &             & 9417        \\ \midrule
\multirow{3}{*}{E. coli}     & Bignorm                    & 1.91       & 26.14      & 67                         & 2279        & 598         \\
                             & Diginorm                   & 17.03      & 19.34      & 63                         & 9105        & 3995        \\
                             & raw                        &            & 24.34      & 64                         &             & 16706       \\ \midrule
\multirow{3}{*}{SAR324}      & Bignorm                    & 4.34       & 33.05      & 55                         & 1222        & 708         \\
                             & Diginorm                   & 4.69       & 23.58      & 52                         & 3706        & 3085        \\
                             & raw                        &            & 32.52      & 51                         &             & 26237       \\ \bottomrule
\end{tabular}

\caption{Filter and assembly statistics for Bignorm with $Q_0=20$, Diginorm and the raw datasets ($\mathrm{I}$)}\label{tab:results1}
\end{center}
\end{table}

\begin{table}
  \begin{center}
    \scriptsize
    \begin{tabular}{@{\ }l@{\ }lr@{\ }r@{\ }r r@{\ }r@{\ }r r@{\ }r@{\ }r rrr@{\ }}\toprule
  \multirow{3}{*}{Dataset}
  & \multirow{3}{*}{Algorithm}
  &\multicolumn{3}{c}{N50}
  &\multicolumn{3}{c}{Longest Contig Length}
  & \multicolumn{3}{c}{Genomic Fraction}
  &\multicolumn{3}{c}{Misassembled Contig Length} \\
  \cmidrule(r){3-5}\cmidrule(r){6-8}\cmidrule(r){9-11}\cmidrule(r){12-14}
  &
  &abs
  &\shortstack{\% of\\raw}
  &\shortstack{\% of\\\scriptsize Diginorm}
  &abs
  &\shortstack{\% of\\raw}
  &\shortstack{\% of\\\scriptsize Diginorm}
  &abs
  &\shortstack{\% of\\raw}
  &\shortstack{\% of\\\scriptsize Diginorm}
  &abs
  &\shortstack{\% of\\raw}
  &\shortstack{\% of\\\scriptsize Diginorm}\\ \midrule
            & Bignorm  & 2324   & 79  & 105	& 11525  & 98  & 100 & 91  & 97  & 97  & 52487   & 148 & 178 \\
  Aceto     & Diginorm & 2216   & 76  &     & 11525  & 98  &     & 94  & 100 &     & 29539   & 84  &     \\
            & raw      & 2935   &     &     & 11772  &     &     & 94  &     &     & 35351   &     &     \\ \midrule
            & Bignorm  & 11750  & 94  & 115 & 43977  & 91  & 95  & 98  & 101 & 105 & 52001   & 120 & 89  \\
Alphaproteo & Diginorm & 10213  & 82  &     & 46295  & 95  &     & 93  & 95  &     & 58184   & 134 &     \\
            & raw      & 12446  &     &     & 48586  &     &     & 98  &     &     & 43388   &     &     \\ \midrule
            & Bignorm  & 3320   & 81  & 97	& 12808  & 57  & 57  & 85  & 100 & 97  & 76797   & 99  & 91  \\
   Arco     & Diginorm & 3434   & 84  &     & 22463  & 100 &     & 88  & 103 &     & 84613   & 109 &     \\
            & raw      & 4092   &     &     & 22439  &     &     & 85  &     &     & 77888   &     &     \\ \midrule
            & Bignorm  & 18432  & 102 & 107	& 108140 & 100 & 100 & 98  & 100 & 100 & 774291  & 91  & 103 \\
   Arma     & Diginorm & 17288  & 96  &     & 108498 & 100 &     & 98  & 100 &     & 748560  & 88  &     \\
            & raw      & 18039  &     &     & 108498 &     &     & 98  &     &     & 849085  &     &     \\ \midrule
            & Bignorm  & 19788  & 91  & 88	& 72685  & 71  & 88  & 97  & 99  & 99  & 2753167 & 94  & 105 \\
  ASZN2     & Diginorm & 16591  & 76  &     & 82687  & 81  &     & 97  & 100 &     & 2617095 & 89  &     \\
            & raw      & 21784  &     &     & 102287 &     &     & 97  &     &     & 2941524 &     &     \\ \midrule
            & Bignorm  & 3356   & 68  & 100	& 25300  & 100 & 100 & 95  & 98  & 99  & 70206   & 105 & 112 \\
Bacteroides & Diginorm & 3356   & 68  &     & 25300  & 100 &     & 96  & 99  &     & 62882   & 94  &     \\
            & raw      & 4930   &     &     & 25299  &     &     & 98  &     &     & 66626   &     &     \\ \midrule
            & Bignorm  & 50973  & 82  & 83	& 143346 & 89  & 91  & 100 & 100 & 100 & 573836  & 94  & 68  \\
  Caldi     & Diginorm & 61108  & 98  &     & 157479 & 98  &     & 100 & 100 &     & 839126  & 138 &     \\
            & raw      & 62429  &     &     & 160851 &     &     & 100 &     &     & 609604  &     &     \\ \midrule
            & Bignorm  & 4515   & 69  & 95	& 20255  & 100 & 107 & 96  & 98  & 98  & 60362   & 86  & 113 \\
  Caulo     & Diginorm & 4729   & 72  &     & 18907  & 93  &     & 98  & 101 &     & 53456   & 76  &     \\
            & raw      & 6562   &     &     & 20255  &     &     & 97  &     &     & 70161   &     &     \\ \midrule
            & Bignorm  & 13418  & 102 & 109	& 79605  & 102 & 102 & 99  & 100 & 100 & 666519  & 95  & 93  \\
Chloroflexi & Diginorm & 12305  & 93  &     & 78276  & 100 &     & 100 & 100 &     & 716473  & 102 &     \\
            & raw      & 13218  &     &     & 78276  &     &     & 99  &     &     & 703171  &     &     \\ \midrule
            & Bignorm  & 6538   & 77  & 91	& 31401  & 81  & 66  & 97  & 99  & 99  & 484354  & 89  & 95  \\
Crenarch    & Diginorm & 7148   & 84  &     & 47803  & 124 &     & 98  & 100 &     & 510256  & 94  &     \\
            & raw      & 8501   &     &     & 38582  &     &     & 98  &     &     & 544763  &     &     \\ \midrule
            & Bignorm  & 5833   & 95  & 99  & 33462  & 98  & 100 & 99  & 101 & 100 & 236391  & 113 & 110 \\
Cyanobact   & Diginorm & 5907   & 96  &     & 33516  & 98  &     & 99  & 101 &     & 214574  & 103 &     \\
            & raw      & 6130   &     &     & 34300  &     &     & 98  &     &     & 209269  &     &     \\ \midrule
            & Bignorm  & 112393 & 100 & 100 & 268306 & 94  & 94  & 96  & 100 & 100 & 28966   & 65  & 65  \\
E. coli     & Diginorm & 112393 & 100 &     & 285311 & 100 &     & 96  & 100 &     & 44465   & 100 &     \\
            & raw      & 112393 &     &     & 285528 &     &     & 96  &     &     & 44366   &     &     \\ \midrule
            & Bignorm  & 135669 & 100 & 114 & 302443 & 100 & 100 & 99  & 100 & 100 & 4259479 & 98  & 100 \\
 SAR324     & Diginorm & 119529 & 88  &     & 302443 & 100 &     & 99  & 100 &     & 4264234 & 98  &     \\
            & raw      & 136176 &     &     & 302442 &     &     & 99  &     &     & 4342602 &     &     \\\bottomrule
\end{tabular}

    \caption{Filter and assembly statistics for Bignorm with $Q_0=20$, Diginorm
      and the raw datasets ($\mathrm{II}$)}\label{tab:results2}
\end{center}
\end{table}

In \autoref{tab:results1}, we give statistics for $Q_0 = 20$ and each case.
In terms of genomic fraction, Bignorm is generally not as good as Diginorm.
However, excluding the Aceto and Arco cases, Bignorm's genomic fraction is still always at least $95\%$.
For Aceto and Arco, Bignorm misses $3.21\%$ and $3.48\%$, respectively, of the genome in comparison to Diginorm.
In 8 cases, Bignorm's N50 is better or at least as good as Diginorm's.
The 4 cases where we have smaller N50 are Arco, Caldi, Caulo, Crenarch, and Cyanobact.

Bignorm's mean phred score is always slightly larger than that of the raw
dataset, whereas Diginorm's is always smaller.  For some cases, the difference
is substantial; the quartiles for the ratio of Diginorm's mean phred score to
that of the raw dataset are given in \autoref{tab:results:quartiles} in the
first row.

Clearly, our biggest gain is in running time, for the filtering as well for the assembly.
Quartiles of the corresponding improvements are given in rows two and three of \autoref{tab:results:quartiles}.

\begin{table}[ht]
  \centering
  \begin{tabular}{lrrrrrr}\toprule
                                          & Min                 & Q1                  & Median              & Mean                & Q3                  & Max                 \\\midrule
    \underline{Diginorm mean phred score} & \multirow{2}{*}{62} & \multirow{2}{*}{66} & \multirow{2}{*}{74} & \multirow{2}{*}{74} & \multirow{2}{*}{79} & \multirow{2}{*}{89} \\
    raw mean phred score                                                                                                                                                      \\\midrule
    \underline{Bignorm filter time}       & \multirow{2}{*}{24} & \multirow{2}{*}{28} & \multirow{2}{*}{31} & \multirow{2}{*}{33} & \multirow{2}{*}{38} & \multirow{2}{*}{46} \\
    Diginorm filter time                                                                                                                                                      \\\midrule
    \underline{Bignorm SPAdes time}       & \multirow{2}{*}{4}  & \multirow{2}{*}{08} & \multirow{2}{*}{18} & \multirow{2}{*}{26} & \multirow{2}{*}{35} & \multirow{2}{*}{88} \\
    Diginorm SPAdes time                                                                                                                                                                  \\\bottomrule
  \end{tabular}
  \caption{Quartiles for comparison of mean phred score, filter and assembly time
    in \%.}\label{tab:results:quartiles}
\end{table}

\section{Discussion}

The quality parameter $Q_0$ that Bignorm introduces over Diginorm has shown to have
a strong impact on the number of reads kept, coverage, and quality of the assembly.
An upper bound of $Q_0 \leq 25$ for a reasonable $Q_0$ was obtained by considering the $10$th percentile of the coverage (\autoref{fig:coverage:2}).
With this constraint in mind, in order to have a small number of reads kept, \autoref{fig:readskept} suggests $18 \leq Q_0 \leq 25$.
Given that N50 for E.coli starts to decline at $Q_0 = 20$ (\autoref{fig:quast}), we decided for $Q_0 = 20$ as the recommended value.
As seen in detail in \autoref{tab:results1}, $Q_0 = 20$ gives good assemblies for all 13 cases.
The gain in speed is considerable: in terms of median we only require $31\%$ and $18\%$
of Diginorm's time for filtering and assembly, respectively.
This speedup generally comes at the price of a smaller genomic fraction and smaller largest contig,
although those differences are relatively small.

\section{Conclusions}

For 13 bacteria single cell datasets, we have shown that good and fast assemblies are possible,
based on only $5\%$ of the reads in most of the cases
(and on less than $10\%$ of the reads in all of the cases).
The filtering process, using our new algorithm Bignorm, also works fast and much faster than Diginorm.
Like Diginorm, we use a count-min sketch for counting \kmers, so our the memory requirements are relatively small and known in advance.
We provide tuning for the quality parameter $Q_0$ and recommend to use $Q_0 = 20$ in practice.
We refrained from tuning the other parameters $c_0$, $c_1$ that are used to define the contributions $b_0(s)$ and $b_1(s)$,
as well as the $\nbN$-count threshold $N_0$ and contribution threshold~$B$.
We expect that tuning of those parameters will help to obtain assemblies of higher quality
and intend to do so in future work.


\end{document}